\documentclass[12pt,english]{article}
\usepackage[T1]{fontenc}
\usepackage{amstext}
\usepackage{graphicx}

\makeatletter

\usepackage[T1]{fontenc}
\usepackage[latin1]{inputenc}
\usepackage{graphics}

\makeatletter

\usepackage[T1]{fontenc}
\usepackage[latin1]{inputenc}

\makeatletter

\usepackage[T1]{fontenc}
\usepackage[latin1]{inputenc}

\makeatletter

\usepackage[T1]{fontenc}
\usepackage[latin1]{inputenc}
\makeatletter
\usepackage{fancyheadings}
\makeatother

\usepackage{babel}
\begin{document}
\begin{flushright}
{\footnotesize 4th Int. Conf. on Multiphase Flow, N. Orleans USA }\\
{\footnotesize May 27-June 1, 2001, paper No. 187} \\

\par\end{flushright}

\vspace{3mm}

\begin{center}
{\Large Particle drift in turbulent flows: the influence of local}\\
{\Large{} structure and inhomogeneity}
\par\end{center}{\Large \par}

\begin{center}
Michael W. Reeks%
\footnote{Present address: School of Mechanical \& Systems Engineering, Stephenson
Building, Newcastle University, Newcastle upn Tyne, NE1 7RU, U.K.;
email: mike.reeks@ncl.ac.uk%
}\\
 {\small Joint Research Centre, European Commission I-21020 Ispra(VA),
Italy} {\small \date{}}
\par\end{center}{\small \par}
\begin{abstract}
The way particles interact with turbulent structures, particularly
in regions of high vorticity and strain rate, has been investigated
in simulations of homogeneous turbulence and in simple flows which
have a periodic or persistent structure e.g. separating flows and
mixing layers. The influence on both settling under gravity and diffusion
has been reported and the divergence (compressibility) of the underlying
particle velocity field along a particle trajectory has been recognized
as an important quantity in quantifying these features. This paper
shows how these features can be incorporated in a formal way into
a two-fluid model of the dispersed particle phase. In particular the
PDF equation for the particle velocity and position is formerly derived
on the basis of a stochastic process that involves the statistics
of both the particle velocity and local compressibility along particle
trajectories. The PDF equation gives rise to contributions to both
the drift and particle diffusion coefficient that depend upon the
correlation of these quantities with the local carrier flow velocity.
\\
\hfill{}\\
\emph{Key Words:} turbulent structures, particle dispersion, drift,
PDF approach 
\end{abstract}
\hfill{}

\begin{centering}

1. INTRODUCTION

\end{centering}

\hfill{}

There are two aspects of the motion of particles in turbulent flows
that have not been properly incorporated in a rational way into a
two-fluid model of a dispersed particle flow, namely the influence
of \emph{persistent structures} in the underlying carrier flow, and
the occurrence of \textit{drift} (either under the influence of gravity
or as a result of inhomogeneity in the underlying turbulence). In
their numerical simulations of particle settling in homogeneous turbulence
and in cellular flow fields, Maxey and his co-workers have shown for
instance that turbulence can enhance the settling of small particles,
(Maxey \& Corrsin 1986, Maxey 1987, Wang \& Maxey 1993). In particular
Maxey (1987) showed that in situations of weak particle inertia (i.e.
particle relaxation times $\ll$ the typical time scale of the turbulent
structures in the flow) the net settling velocity $\mathbf{V}_{g}$
of an ensemble of particles in a homogeneous flow field was related
to its value $\mathbf{V}_{g}^{0}$ in quiescent flow by the relationship
\begin{equation}
\mathbf{V}_{g}=\mathbf{V}_{g}^{0}-\int_{0}^{t}\left\langle \mathbf{u}\left(\mathbf{x},t\right)\nabla\cdot\mathbf{v}_{p}\left(\mathbf{X}_{p}(\mathbf{x},t\mid s),s\right)\right\rangle ds,\label{driftvelocity}
\end{equation}
 where $\left\langle .....\right\rangle $ is an ensemble average;
$\mathbf{u}\left(\mathbf{x},t\right)$ is the carrier flow turbulent
velocity field at position $\mathbf{x}$ for times $t\gg$ integral
time scale of the turbulent motion; $\nabla\cdot\mathbf{v}_{p}\left(\mathbf{X}_{p}(\mathbf{x},t\mid s),s\right)$
is the divergence of the particle velocity field $\mathbf{v}_{p}\left(\mathbf{y},s\right)$
with respect to the spatial position $\mathbf{y}$ measured at $\mathbf{y}=\mathbf{X}_{p}(\mathbf{x},t\mid s)$
at time $s$ where $\mathbf{X}_{p}(\mathbf{x},t|s)$ is the position
of a particle at $s$ which arrives at $\mathbf{x}$ at time $t.$
The particle velocity field $\mathbf{v}_{p}\left(\mathbf{y},t\right)$
is defined as the particle velocity field arising from one realisation
of the flow field $\mathbf{u}\left(\mathbf{y},t\right)$ with a prescribed
set of initial conditions at $s=0$ for the particles which are the
same in each realisation of the flow field.

The divergence of the particle velocity field is a measure of the
local compressibility of the particle flow. The presence of gravity
means that the particles move in a preferential direction which in
turn means that the correlation of the fluid velocity with the `local'
divergence of the particle flow field is non-zero. For the case when
the particles almost followed the flow, Maxey was able to relate the
local compressibility of the particle flow field to the local straining
of the underlying carrier flow and showed that the value of the correlation
would lead to an enhancement of the gravitational settling. Subsequently
Wang \& Maxey(1987) explained this result in more detail by looking
at the way particles move around the edges of vortices; in particular
their results could be explained by the streaming of particles between
vortices which always lead to an accumulation of particles on the
down flow side of vortices (i.e. in the direction of gravity). They
referred to this process as \textit{preferential sweeping}. This however
is not a unique result. For instance depending upon the particle \emph{Froude
number}, Davila and Hunt (1999) have shown it is possible for the
opposite to occur.

Whatever the particular route the particles take through a flow field
(with or without gravity), the compressibility of the particle flow
field measured along a particle trajectory is an important consideration
in the way we assess the influence of structures. The work presented
here shows that Maxey's expression for the drift is a much more general
result appropriate for inhomogeneous as well as homogeneous flows
with or without the presence of gravity. Indeed it is shown that the
compressibility of the particle flow can influence not only the drift
but also the particle dispersion. As a prelude to the full two- fluid
formulation I first consider in Section 2 the analysis of particle
dispersion and drift in a compressible flow field in which the statistics
of both the particle velocity and the divergence of the particle flow
field along a particle trajectory are prescribed and correlated. Then
finally in Section 3 these features are incorporated into a two-fluid
model of the dispersed particle phase, based on the so-called \emph{pdf
approach} - the focus here being on the derivation of the appropriate
transport equation for the particle phase space probability that a
particle has a velocity \textbf{v} and position \textbf{x} at time
$t.$ Some of the features of this analysis are illustrated in particle
dispersion in a random array of counter-rotating vortices in both
homogeneous and inhomogeneous situations depending upon the prescribed
statistics. 

\hfill{}

\lhead{ \large Particle drift in turbulent flows } \rhead{M. W. Reeks}

\begin{centering}

2. PASSIVE\ SCALAR DISPERSION\ IN A\ COMPRESSIBLE\ FLOW\label{sec: 2}

\end{centering} \hfill{}\\
 \emph{2.1 Gaussian and non-Gaussian Lagrangian Statistics}\vspace{1.5mm}\\
 Analyses of this sort have been done before for passive scalar diffusion
in incompressible flow in which case only the statistics of the particle
velocity along a particle trajectory are required. In the case of
a compressible flow, moments associated with the process $\left[{\mathbf{v}}_{p}\left(s\right),{\nabla\cdot\mathbf{v}}_{p}\left(s\right),s\in t\right]$
appear as a natural consequence of the transport and the compressibility
of the flow, where both the particle velocity and the divergence of
the particle velocity fields are measured along a particle trajectory.

The starting point of the analysis is the conservation equation for
the particle mass density $\rho(\mathbf{x},t)$ at position $\mathbf{x}=[x_{1},x_{2},x_{3}]$
at time $t$, namely 
\begin{eqnarray}
\frac{\partial\rho}{\partial t} & = & -\nabla\cdot\left\{ \mathbf{v}_{p}\rho(\mathbf{x},t)\right\} ,\nonumber \\
\textrm{or }\frac{D\rho}{Dt} & = & -\rho\,\nabla\cdot\mathbf{v}_{p}(\mathbf{x},t).\label{particledensityeqn}
\end{eqnarray}
 where is $D/Dt$ rate of change along a particle trajectory. Given
some initial distribution $\rho(\mathbf{x},t_{0})$ at time $t=0$,
the solution is formerly
\begin{equation}
\rho(\mathbf{x},t)=\rho\left(\mathbf{X}_{p}(\mathbf{x},t|0),0\right)\exp\left\{ -\int_{t_{0}}^{t}\nabla\cdot\mathbf{v}_{p}\left(\mathbf{X}_{p}(\mathbf{x},t|s),s\right)\, ds\right\} .\label{eq1forrho}
\end{equation}
 where $\mathbf{X}_{p}(\mathbf{x},t|s)$ is the position at time $s$
of a particle arriving at $\mathbf{x}$ at time $t$. Given that in
principle we can define a particle velocity field $\mathbf{v}_{p}(\mathbf{x},t)$
for any realization of the underlying carrier flow field, then the
problem of particle dispersion and settling is identical to the problem
of passive scalar dispersion in a velocity field differing only from
the normal case considered in that the particle velocity field is
compressible rather than solenoidal. Replacing $\mathbf{X}_{p}(\mathbf{x},t|0)$
by $\mathbf{x}-\int_{0}^{t}\mathbf{v}_{p}\left(\mathbf{X}_{p}(\mathbf{x},t|s),s\right)\, ds$
in Eq.(\ref{eq1forrho}) we obtain \-
\begin{equation}
\rho(\mathbf{x},t)=\rho\left(\mathbf{x}-\int_{0}^{t}\mathbf{v}_{p}(\mathbf{x},t\mid s)\, ds,0\right)\exp\left\{ -\int_{0}^{t}\nabla\cdot\mathbf{v}_{p}(\mathbf{x},t\mid s)\, ds\right\} .\label{eq.2forrho}
\end{equation}
 where $\mathbf{v}_{p}(\mathbf{x},t\mid s)$ and $\nabla\cdot\mathbf{v}_{p}(\mathbf{x},t\mid s)$
are used as shorthand notation for the explicit values of the particle
velocity and divergence along particle trajectories that pass through
$(\mathbf{x},t)$,%
\footnote{It is implicit here that the divergence be applied to the spatial
components of the particle velocity field and is not meant to operate
on $\mathbf{x}$ in the vector function $\mathbf{v}_{p}(\mathbf{x},t\mid s)$
.%
} namely 
\begin{equation}
\mathbf{v}_{p}(\mathbf{x},t\mid s)\equiv\mathbf{v}_{p}\left(\mathbf{X}_{p}(\mathbf{x},t|s),s\right)\,\,\,\,\,\,\,\,\nabla\cdot\mathbf{v}_{p}(\mathbf{x},t\mid s)\equiv\nabla\cdot\left.\mathbf{v}_{p}(\mathbf{y},s)\right]_{\mathbf{y}=\mathbf{X}_{p}(\mathbf{x},t|s)}
\end{equation}
 We shall sometimes abbreviate these quantities still further to $\mathbf{v}_{p}(s)$
and $\left[\nabla\cdot\mathbf{v}_{p}\right](s)$ respectively. By
making certain assumptions about the statistics of the process $\left[\mathbf{v}_{p}(s),\nabla\cdot\mathbf{v}_{p}\left(s\right)\right]$
for $0\leq s\leq t$ , then we are avoiding the non-linearity of the
diffusion process that is implicit in the relationship between Lagrangian
and Eulerian timescales. As a result it is shown in Appendix A that
if this process is jointly Gaussian then the particle drift velocity
is given precisely by the term in Eq(\ref{driftvelocity}) and the
diffusion coefficient consistent with Taylor's theory. Explicitly
the particle mass current is given by: 
\begin{eqnarray}
\left\langle \rho\mathbf{v}_{p}(\mathbf{x},t)\right\rangle  & = & \left\{ \left\langle \mathbf{v}_{p}\left(\mathbf{x},t\right)\right\rangle -\int_{0}^{t}\left\langle \mathbf{v}_{p}^{\prime}\left(\mathbf{x},t\right)\,\nabla\cdot\mathbf{v}_{p}(\mathbf{x},t\mid s)\right\rangle ds\right\} \left\langle \rho(\mathbf{x},t)\right\rangle \nonumber \\
 &  & -\int_{0}^{t}ds\left\langle \mathbf{v}_{p}^{\prime}\left(\mathbf{x},t\right)\,\mathbf{v}_{p}^{\prime}\left(\mathbf{x},t\mid s\right)\right\rangle \cdot\nabla\left\langle \rho(\mathbf{x},t)\right\rangle 
\end{eqnarray}
 where $\mathbf{v}_{p}^{\prime}\left(\mathbf{x},t\right)$ is the
fluctuating part of $\mathbf{v}_{p}\left(\mathbf{x},t\right)$ relative
to its mean. The first bracketed term on the RHS \textit{\emph{(the
drift term)}} in this equation is identical to the drift term derived
by Maxey if we substitute $\mathbf{V}_{g}^{0}$for $\left\langle \mathbf{v}_{p}(\mathbf{x},t)\right\rangle $.
However with the assumption that the statistics for the underlying
particle velocity field are Gaussian, we end up with a more general
result which includes a gradient diffusion flux. If in general the
statistics of the process $\left[\mathbf{v}_{p}(t),\nabla\cdot\mathbf{v}_{p}\left(t\right)\right]$
are non-Gaussian then it is shown in Appendix A to first order in
the triple moments of the process, that the particle mass current
is compounded of a drift term
\begin{eqnarray}
\mathbf{v}_{d} & = & \left\langle \mathbf{v}_{p}(\mathbf{x},t)\right\rangle -\int_{0}^{t}ds_{1}\left\langle \nabla\cdot\mathbf{v}_{p}(\mathbf{x},t\mid s_{1})\mathbf{v}_{p}^{\prime}(\mathbf{x},t)\right\rangle +\nonumber \\
 &  & +\frac{1}{2}\int_{0}^{t}ds_{1}\int_{0}^{t}ds_{2}\left\langle \mathbf{v}_{p}^{\prime}(\mathbf{x},t)\nabla\cdot\mathbf{v}_{p}(\mathbf{x},t\mid s_{1})\nabla\cdot\mathbf{v}_{p}(\mathbf{x},t\mid s_{2})\right\rangle \label{eq:nonGaussiandriftvelocity}
\end{eqnarray}

and a gradient diffusion term with diffusion coefficients $D_{ij}$\emph{
\begin{equation}
D_{ij}=\int_{0}^{t}ds_{1}\left\langle \textrm{v}_{p_{i}}^{\prime}(\mathbf{x},t\mid s_{1})\textrm{v}_{p_{j}}^{\prime}(\mathbf{x},t)\right\rangle -\int_{0}^{t}ds_{1}\int_{0}^{t}ds_{2}\,\left\langle \textrm{v}_{p_{j}}(\mathbf{x},t)\textrm{v}_{p_{i}}^{\prime}(\mathbf{x},s_{2})\nabla\cdot\mathbf{v}_{p}(\mathbf{x},t\mid s_{1})\right\rangle \label{DijinnonGaussianfields}
\end{equation}
}

\emph{2.2 Comments on the process $[\mathbf{v}_{p}(s),\nabla\cdot\mathbf{v}_{p}\,;\,0\leq s\leq t]$} \vspace{1.5mm}\\
 It is important to recognize that the statistics of the process $[\mathbf{v}_{p}(s),\nabla\cdot\mathbf{v}_{p}(s)\,;\,0\leq s\leq t]$
which we will call $[\mathbf{q}(s)]$ for short, does not depend on
the initial concentration. If it did, then its statistics would be
related to the particle density weighted averages we are trying to
calculate in the first place . How these statistics are obtained is
clear: a particle trajectory is soved backwards in time starting from
$\mathbf{x}$ at time $t$ using the values of the particle velocity
$\mathbf{v}_{p}(s)$ along its trajectory which in turn are derived
from the statistics of the velocity field $\mathbf{v}_{p}(\mathbf{x},s)$
where $0\leq s\leq t$. No restrictions are placed on the point the
trajectory goes through at time zero. In the actual problem of interest
we might want to know the average particle velocity at $\mathbf{x}$
at time $t$ knowing say that the particles started out at $\mathbf{x}_{0}$
at time zero. So these particular particles will choose a particular
subset of the statistics of the process $\left[{\mathbf{q}}(s)\right]$
in arriving at ${\mathbf{x}}$ at time $t.$ That is, we are selecting
only those trajectories of all those trajectories defined by the process
$\left[{\mathbf{q}}(s)\right]$ that go through $\mathbf{x}_{0}$
at time zero from which we could compute the particles average velocity
at $\mathbf{x}$ at time $t$. Put another way, we are trying to evaluate
the particle statistics from a set of statistics which are independent
of where the particles start from in the actual problem of interest.
You can see this more transparently in the way the concentration is
calculated. You start off with some prescribed statistics for the
process $\left[{\mathbf{q}}(s)\right]$ found by starting a test particle
off at ${\mathbf{x}}$ and solving the equation of motion backwards
in time from $t$. That is you solve 
\begin{equation}
\frac{d{\mathbf{X}_{p}}}{ds}={\mathbf{v}}_{p}({\mathbf{X}_{p}},s){\,\,\textrm{with}\,{\mathbf{X}_{p}(t)}={\mathbf{x}}}
\end{equation}
 backwards in time to find the values of ${\mathbf{X}_{p}}(0)$and
the value of exponential of the integral of the value ${\nabla\cdot\mathbf{v}}_{p}$
along a trajectory (which gives the fractional change in an elemental
volumee at time $t$ along the trajectory relative to its initial
value i.e. the value of the elemental volume deformation $J(t)=\left|\frac{\partial\mathbf{X}_{p}(\omega,\mathbf{y},t^{\prime}|t)}{\partial\mathbf{y}}\right|$.
You then calculate the concentration that particles would have at
$\left[{\mathbf{x}},t\right]$ if they started out at time $0$ with
some concentration $\rho({\mathbf{X}_{p}}(0),0)$ by multiplying this
concentration by $J(t)$. If the concentration at ${\mathbf{X}_{p}}(0)$
happens to be zero, then the concentration at ${\mathbf{x}}$ is zero.
The fact that there may not be any particles at $\mathbf{X}_{p}(0)$
doesn't affect the statistics of the process $\left[{\mathbf{q}}(s)\right]$.
The process just tells you where you might find some particles at
time $0$ but if there aren't any, then that's because of the initial
conditions.The process $\left[{\mathbf{q}}(s)\right]$ doesn't know
about initial conditions or concentration. It's entirely determined
from the statistics of ${\mathbf{v}}_{p}({\mathbf{x}},t)$ derived
from some test particle at ${\mathbf{x}}$ at time $t$ in the way
we have prescribed. \vspace{3.0mm}\\
 \emph{2.2 Dispersion in homogeneous staionary turbulence and comparison
with Taylor's Theory} \vspace{1.5mm}\\
 It is revealing to compare these results derived for passive scalar
diffusion in a compressible flow field with G I Taylor's classic theory
for diffusion by continuous movements in a homogneous stationary flow
field? We recall that Taylor's results are based on the assumption
that in the limit of the dispersion time $t\rightarrow\infty$ this
time can be divided into a large number of time steps (each step >\textcompwordmark{}>
the integral timescale of the turbulence) so that the distance travelled
in one time step will be uncorrelated with the distance travelled
in the next. This leads to Gaussian statistics for the particles displacement.
In particular the diffusion coefficient is written as 
\begin{equation}
D_{ij}=\int_{0}^{\infty}R_{ij}(s)ds,
\end{equation}
 where $\,\, R_{ij}(s)$ defines the velocity autocorrelation $\left\langle \textrm{v}_{p_{i}}(\mathbf{x},0\mid s_{1})\textrm{v}_{p_{j}}(\mathbf{x},0\mid s_{2})\right\rangle $
for which $s=s_{1}-s_{2}$, and $\textrm{v}_{p_{i}}(\mathbf{x},0\mid s_{1})$
and $\textrm{v}_{p_{j}}(\mathbf{x},0\mid s_{2})$ as before are the
velocities of a particle measured at time $s_{1}$ and $s_{2}$ that
particle starting out at some arbitrary position ${\mathbf{x}}$ at
some arbitrary time $t=0.$ The important requirement is the distances
at which these measurements take place are on average very far away
from the point of release ${\mathbf{x}}$ so that the process $[\mathbf{v}_{p}(\mathbf{x},0\mid s)]$
is stationary i.e for this to occur $s/T_{L}\gg1$ where $T_{L}$
is the Lagrangian integral timescale. The rate of change of the mean
square displacement is given by 
\begin{equation}
\left|\frac{d}{dt}\left\langle X_{p_{i}}({\mathbf{x}},0|t)^{2}\right\rangle \right]=2\left\langle \textrm{v}_{p_{i}}(\mathbf{x},0\mid t){X_{p_{i}}}({\mathbf{x}},0|t)\right\rangle =2\int_{0}^{t}\left\langle \textrm{v}_{p_{i}}(\mathbf{x},0\mid t)\textrm{v}_{p_{i}}(\mathbf{x},0\mid s)\right\rangle ds\label{dy2/dt}
\end{equation}
 and with the assumption that one can replace the lower limit \textbf{$0$}
by say $\tau$ such that $t$ $\gg t-\tau\gg T_{L}$, so that during
the interval $\,\,\tau\leq s\leq t,\mathbf{v}_{p_{i}}(\mathbf{x},0\mid s)$
is stationary, one arrives at the classic result that 
\begin{equation}
\frac{1}{2}\left|\frac{d}{dt}\left\langle X_{p_{i}}({\mathbf{x}},0|t)^{2}\right\rangle \right]_{t\rightarrow\infty}=\int_{0}^{\infty}R_{ij}(s)ds\label{Taylorsresult}
\end{equation}
 a result which is self consistent with a Gaussian or gradient diffusion
process with a diffusion coefficient given by the RHS of Eq(\ref{Taylorsresult}).
Returning to the form for the diffusion coefficient defined in Eq.()
that is the extension to non-Gaussian fields, we note that with the
Taylor Gaussian assumption for $t\rightarrow\infty$,  we are left
with the result that 
\begin{equation}
D_{ij}=\int_{0}^{t}ds_{1}\left\langle \textrm{v}_{p_{i}}^{\prime}(\mathbf{x},t\mid s)\textrm{v}_{p_{j}}^{\prime}(\mathbf{x},t)\right\rangle 
\end{equation}
 which since $\left\langle \textrm{v}_{p_{i}}^{\prime}(\mathbf{x},t\mid s)\textrm{v}_{p_{j}}^{\prime}(\mathbf{x},t)\right\rangle $
is dependent only on $t-s$ we get the same result as in Eq.(\ref{Taylorsresult})
if 
\begin{equation}
\left\langle \textrm{v}_{p_{i}}^{\prime}(\mathbf{x},t\mid s)\textrm{v}_{p_{j}}^{\prime}(\mathbf{x},t)\right\rangle =\left\langle \textrm{v}_{p_{i}}^{\prime}(\mathbf{x},0\mid s)\textrm{v}_{p_{j}}^{\prime}(\mathbf{x},0\mid t)\right\rangle \label{eq:correlationidentity}
\end{equation}

If the flow is homogeneous and stationary then these correlations
since they refreed to the same particle measured at two different
times, will be independent of labeling position and times. That is
we could change the labeling time from $t$ to $0$ in the correlation
on the RHS of Eq.(\ref{eq:correlationidentity}) and retain the labeling
position $\mathbf{x}$ without changing the result. So the relationship
in Eq.(\ref{eq:correlationidentity}) is valid. The relevance of this
independednce on labelling times and positions is even more revealing
when we consider the general result for the rate of mean square displacement
given in Eq.(\ref{dy2/dt}) for all $t$ and compare it with the form
derived from the continuity equation
\[
\frac{\partial\rho}{\partial t}=\frac{\partial}{\partial\mathbf{x}}\cdot\langle\rho\mathbf{v}_{p}\rangle
\]
 on the form for $D_{ij}$ in Eq.(\ref{DijinnonGaussianfields}) appropriate
for non-Gaussian fields. That is if we release particles at time $0$
and measure the dispersion at time $t$ then the form of $D_{ij}$
would imply that 
\begin{equation}
\left\langle \textrm{v}_{p_{i}}[\mathbf{X}_{p}(\mathbf{x},0|t),t]X_{p_{i}}({\mathbf{x}},0|t)\right\rangle =\left\langle \textrm{v}_{p_{i}}(\mathbf{x},t)X_{p_{i}}(\mathbf{x},t|0)\right\rangle -\int_{0}^{t}ds_{1}\int_{0}^{t}ds_{2}\left\langle \nabla\cdot\mathbf{v}_{p}(s_{1})\,\textrm{v}_{p_{i}}^{\prime}(s_{2})\textrm{v}_{p_{j}}^{\prime}(t)\right\rangle +....
\end{equation}
 The statistics associated with the correlation on the LHS of this
equation is different from that determining the first term on the
RHS of the equation: in LHS case, we have statistics derived from
two Lagrangian variables where arguments for stationarity can only
be invoked when $t\rightarrow\infty$, whilst the case of the RHS
is derived from a Lagrangian and an Eulerian variable. The two are
only equal when $t\rightarrow\infty$ or at small times $t\ll T_{L.}$
when the second term on the RHS is $O(t/T_{L})$ smaller. The term
on the RHS is clearly a measure of the difference in the two sorts
of statistics.

\hfill{}

\begin{centering}

3. PDF FORMULATION

\end{centering} \hfill{}

This represents an extension of previous work by this author (Reeks
1991, 1992) and several others (Zaichik 1991, Swailes 1997, Hyland
et al. 1999, Pozorski \& Minier 1999 and Simonin et al. 1999) in using
an equation for the particle phase space probability to formally derive
the two-fluid continuum equations for the particle phase.\vspace{1.5mm}\\
 \emph{2.1} \emph{Definition and prescription of particle velocity
field} \emph{and its divergence}\vspace{1.5mm}

Using Stokes drag as an example, the particle equation of motion can
be written as
\begin{equation}
\frac{d\mathbf{v}}{dt}=\beta\{\mathbf{u}(\mathbf{x},t)-\mathbf{v}\}+\mathbf{g}\,\,\,\,;\,\,\,\:\frac{d\mathbf{x}}{dt}=\mathbf{v}\label{Stokesdrageqnofmotion}
\end{equation}
 where as before $\mathbf{u}(\mathbf{x}_{p},t)$ is the underlying
carrier flow velocity at position $\mathbf{x}$ at time $t.$ The
solution can be written in several ways. First solving the set as
a time problem,\textbf{
\begin{equation}
\mathbf{v}=\mathbf{V}_{p}(\omega,\mathbf{y},t\prime\mid t)\,\,\,\,;\,\,\,\,\mathbf{x}=\mathbf{X}_{p}(\omega,\mathbf{y},t^{\prime}\mid t)\label{X_{p}(w,y,t'/t)...}
\end{equation}
} i..e. the solution is the particle velocity/position at time $t$,
for a particle with initial velocity $\omega$ and position \textbf{$\mathbf{y}$}
at time $t^{\prime}$, allowing for the possibility of $t^{\prime}$
being in the past or the future in relation to $t$. Clearly these
functions define the inverse relations
\begin{equation}
\omega=\mathbf{V}_{p}(\mathbf{v},\mathbf{x},t\mid t^{\prime})\,\,\,\,;\,\,\,\,\mathbf{y}=\mathbf{X}_{p}(\mathbf{v},\mathbf{x},t\mid t^{\prime})
\end{equation}
 So using these equations we could eliminate $\mathbf{y}$ from Eq.
(\ref{X_{p}(w,y,t'/t)...}) and write an alternative solution, namely
\begin{equation}
\mathbf{v}=\mathbf{v}_{p}(\omega,\mathbf{x},t^{\prime}\mid t).\label{particevelocityfield}
\end{equation}
 That is the particle velocity at position \textbf{x} at time $t$
given that the particle velocity started out at time \textbf{$t^{\prime}$}with
a velocity \textbf{$\omega$. $\mathbf{v}_{p}(\omega,\mathbf{x},t^{\prime}\mid t)$}
is the particle velocity field (in the context of the passive scalar
dispersion in Section 2), which satisfies the equation
\begin{equation}
\frac{d\mathbf{X}_{p}}{dt}=\mathbf{v}_{p}(\omega,\mathbf{X}_{p},t^{\prime}|t).
\end{equation}
 So we have 
\begin{equation}
\frac{d}{dt}\left\{ \frac{\partial X_{p_{i}}(\omega,\mathbf{y},t^{\prime}|t)}{\partial y_{j}}\right\} =\left(\frac{\partial\textrm{v}_{p_{i}}}{\partial X_{p_{j}}}\right)\left(\frac{\partial X_{p_{i}}}{\partial y_{j}}\right).\label{DE.forthedx/dy}
\end{equation}
 Given that the initial conditions imply that 
\begin{equation}
\frac{\partial X_{p_{i}}}{\partial y_{j}}=\delta_{ij}\,\,\,\textrm{at}\, t=t^{\prime},
\end{equation}
 then we would have directly from Eq.(\ref{DE.forthedx/dy}) that
\begin{equation}
J=\left|\frac{\partial\mathbf{X}_{p}(\omega,\mathbf{y},t^{\prime}|t)}{\partial\mathbf{y}}\right|=exp\left\{ \int_{t^{\prime}}^{t}ds\,\left.\nabla\cdot\mathbf{v}_{p}(\omega,\mathbf{y},t^{\prime}|s)\right|_{\mathbf{y}=\mathbf{X}_{p}(\mathbf{x},t|s)}\right\} 
\end{equation}
 
\begin{figure}
\begin{centering}
\includegraphics[width=0.6\textwidth]{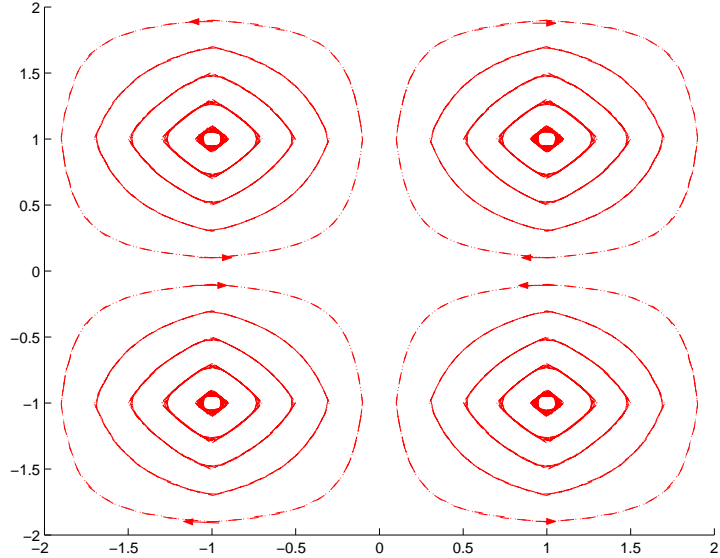}
\par\end{centering}

\caption{\label{pairs of c-r vortices}Pairs of counter-rotating vortices generated
from random symmetric shear flow }
\end{figure}
\emph{}\\
\emph{2.2} \emph{Closure of the PDF Equation}\vspace{1.5mm}

If $W({\mathbf{x}},{\mathbf{v}},t)$ is the phase space density for
a particle with velocity \textbf{v} and position $\mathbf{x}$ at
time $t$ subject to the equation of motion defined in Eq.(\ref{Stokesdrageqnofmotion})
for one realisation of the carrier flow filed $\mathbf{u}(\mathbf{x},t)$,
then the equation f\textbf{or $\langle W(\mathbf{v},\mathbf{x},t)\rangle$}
the PDF for a particle to have $(\mathbf{v},\mathbf{x},t)$ is obtained
by averaging the Liouville equation thus,
\begin{equation}
\left[\frac{\partial}{\partial t}+\mathbf{v}\cdot\frac{\partial}{\partial\mathbf{x}}+\frac{\partial}{\partial\mathbf{v}}\cdot\beta\{\langle\mathbf{u}(\mathbf{x},t)\rangle-\mathbf{v}\}\right]\langle W\rangle=-\frac{\partial}{\partial\mathbf{v}}\cdot\beta\langle\mathbf{u}^{\prime}(\mathbf{x},t)W\rangle\label{pdfeq.}
\end{equation}
 where $\langle\mathbf{u}(\mathbf{x},t)\rangle$ and $\mathbf{u}^{\prime}(\mathbf{x},t)$
are the mean and fluctuating components of $\mathbf{u}(\mathbf{x},t)$.
We require therefore a closed expression for \textbf{$\langle\mathbf{u}^{\prime}(\mathbf{x},t)W\rangle$}.
In reality we consider a closed expression for the specific case when
$W$ is a response function $G$, that is it is the solution for an
instantaneous point source $\delta(\mathbf{v}-\mathbf{v}^{\prime})\delta(\mathbf{x}-\mathbf{x}^{\prime})\delta(t-t^{\prime})$.
Thus $\langle G\rangle$ is the solution of the PDF equation Eq.(\ref{pdfeq.})
with the instantaneous point source added to the RHS of the equation.
Knowing $\langle G\rangle$ we have for \textbf{$\langle W(\mathbf{v},\mathbf{x},t)\rangle$
\begin{equation}
\langle W(\mathbf{v},\mathbf{x},t)\rangle=\int d\mathbf{v}\, d\mathbf{x}\langle G(\mathbf{v},\mathbf{x},t|\mathbf{v}^{\prime},\mathbf{x}^{\prime},t^{'})\rangle\rho(\mathbf{v}^{\prime},\mathbf{x}^{\prime},t^{\prime})d\mathbf{v}^{\prime}d\mathbf{x}^{\prime}
\end{equation}
} where $\rho(\mathbf{v}^{\prime},\mathbf{x}^{\prime},t^{\prime})$
is some initial distribution of $\langle W\rangle$ at time $t^{\prime}$.
With reference to Eqs.(\ref{X_{p}(w,y,t'/t)...}), we can write the
solution for $\langle G\rangle$ formally as
\begin{equation}
\langle G\rangle=\left\langle \delta(\mathbf{v}-\mathbf{V}_{p}(\mathbf{v}^{\prime},\mathbf{x}^{\prime},t^{\prime}\mid t))\delta(\mathbf{x}-\mathbf{X}_{p}(\mathbf{v}^{\prime},\mathbf{x}^{\prime},t^{\prime}\mid t)\right\rangle 
\end{equation}
 However using the definition of the particle velocity field $\mathbf{v}_{p}(\mathbf{v}^{\prime},t^{\prime}|\mathbf{x},t$)
we can write this alternatively as
\begin{eqnarray}
\langle G\rangle=\langle\delta(\mathbf{v}-\mathbf{v}_{p}(\mathbf{v}^{\prime},t^{\prime}\mid\mathbf{x},t))\delta(\mathbf{x}-\int_{t^{\prime}}^{\textrm{t}}ds\,\mathbf{v}_{p}(\mathbf{v}^{\prime},t^{\prime\,}|\,\mathbf{y},s)-\mathbf{x}^{\prime})\,\nonumber \\
exp\{-\left.\int_{t^{\prime}}^{t}ds\,\nabla\cdot\mathbf{v}_{p}(\mathbf{v}^{\prime},t^{\prime}\,|\,\mathbf{y},s)\right|_{\mathbf{y}=\mathbf{X}(\mathbf{x},t|s)}\}\rangle\label{<G>equation}
\end{eqnarray}

Similarly we can write down formally an expression for $\langle G\mathbf{u}\rangle$.
This expression together with that for $\langle G\rangle$ are in
form that we can process in a similar manner to the evaluation of
$\langle\rho\mathbf{v}_{p}\rangle$ for the passive scalar case: the
only difference here is we are considering a process $\left[\mathbf{v}_{p}(s),\nabla\cdot\mathbf{v}_{p}(s),\mathbf{u}(s)\,;\, t^{\prime}\leq s\leq t\right]$
as opposed to $\left[\mathbf{v}_{p}(s),\nabla\cdot\mathbf{v}_{p}(s)\,;\, t^{\prime}\leq s\leq t\right]$.
We show in Appendix B that if this process is Gaussian, then $\beta\langle G\mathbf{u}\rangle$
is given exactly by
\begin{equation}
\beta\langle\mathbf{u}(\mathbf{x},t)G(\mathbf{v},\mathbf{x},t|\mathbf{v}^{\prime},\mathbf{x}^{\prime},t^{\prime}\rangle=-\left(\underline{\mu}\cdot\frac{\partial}{\partial\mathbf{v}}+\underline{\lambda}\cdot\frac{\partial}{\partial\mathbf{x}}\right)\langle G\rangle+\underline{\gamma}\langle G\rangle\label{Phasespacediffusioncurrent}
\end{equation}
 where
\begin{eqnarray}
\underline{\mu} & = & \beta\left\langle \mathbf{u}^{\prime}(\mathbf{x},t)\mathbf{v}_{p}(t)\right\rangle \nonumber \\
\underline{\lambda} & = & \beta\left\langle \mathbf{u}^{\prime}(\mathbf{x},t)\mathbf{x}_{p}(t)\right\rangle \nonumber \\
\underline{\gamma} & = & -\beta\int_{t^{\prime}}^{t}ds\,\left\langle \mathbf{u}^{\prime}(\mathbf{x},t)\nabla\cdot\mathbf{v}_{p}(s)\right\rangle ,
\end{eqnarray}
 where 
\begin{eqnarray}
\mathbf{v}_{p}(s)\equiv\mathbf{v}_{p}(\mathbf{v}^{\prime},t^{\prime}|\mathbf{X}_{p}(\mathbf{x},t|s),s) &  & \nabla\cdot\mathbf{v}_{p}(s)\equiv\nabla\cdot\left.\mathbf{v}_{p}(\mathbf{v}^{\prime},t^{\prime}|\mathbf{y},s)\right|_{\mathbf{y}=\mathbf{X}_{p}(\mathbf{x},t|s)}\\
\mathbf{x}_{p}(t) & \equiv & \mathbf{X}_{p}(\mathbf{x},t|0)=\int_{0}^{t}\mathbf{v}_{p}(\mathbf{v}^{\prime},t^{\prime}|\mathbf{X}_{p}(\mathbf{x},t|s),s)ds
\end{eqnarray}

The form of the net force per unit mass of particles due to the turbulence
given in Eq.(\ref{Phasespacediffusioncurrent}) is therefore composed
of two parts: a \emph{diffusive force} (gradient of a stress tensor)
which depend upon gradients in the particle velocity and position
{[}the bracketed term on the RHS of Eq.(\ref{Phasespacediffusioncurrent}){]}
and a \emph{body force} which depends upon the local compressibility
of instantaneous particle velocity field along a particle trajectory
{[}the second term in Eq.(\ref{Phasespacediffusioncurrent}){]}. The
general form of this \emph{turbulent force} has been obtained before
by several authors (Reeks 1992, Swailes et. al. 1997, Pozorski and
Minier 1999, Hyland et. al. 1999) but the precise form for the body
force is different from the one derived here and leads to the so-called
problem of spurious drift; that is there is a drift term that persists
in cases where the particles follow the underlying incompressible
flow ($\beta^{-1}\rightarrow0)$ so that where at equilibrium the
particles ought to be fully mixed with the flow, the existence of
the spurious drift leads to a build up of concentration in regions
of low turbulence intensity. It is a feature that is common in certain
types of simple random walk simulation of particle dispersion in inhomogeneous
turbulence (where the underlying flow filed is essentially 1-D and
cannot of its own accord satisfy continuity of flow if it is spatially
varying. The form derived here does not suffer from this serious defect,
the drift velocity $\mathbf{v}_{d}$ in this case $\beta^{-1}\underline{\gamma}$,
clearly vanishes when the particle follow the flow, because $\nabla\cdot\mathbf{v}_{p}(s)$
is the same as that of the underlying carrier flow which is necessarily
zero.

As an illustration of the influence of turbulent structures I have
considered the dispersion of particles in a random flow field which
consists of pairs of counter-rotating vortices (see Fig.1) with randomly
generated vorticity that shifts randomly in position as the timescale
of the vorticity changes randomly from one value to the next in time.
In the case of a flow field in which the location and periodicity
of the structures is fixed, particles accumulate at the stagnation
points. That is the process is equivalent to diffusion plus a drift
directed towards the stagnation point. As an example Fig.2 shows the
difference in behaviour between a 1 D flow field in which the particles
are constrained to move only in the x-direction and when they allowed
to move in the y -direction (fully 2D vortex flow field). The difference
illustrates the difference between spurious drift (arising from a
1 D carrier flow field which is compressible) and the case of an incompressible
2D carrier flow.
\begin{figure}
\begin{centering}
\includegraphics[width=0.6\textwidth]{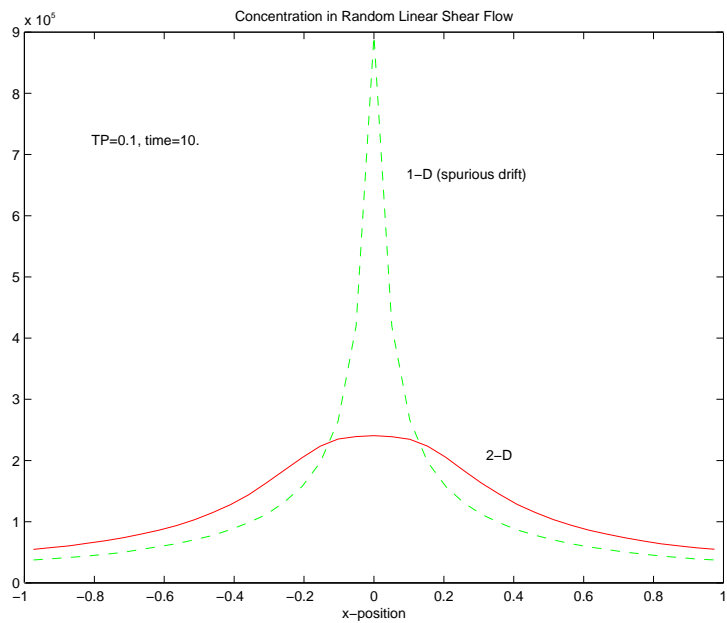}
\par\end{centering}

\caption{Particle concentration profiles in random pairs of counter-rotating
vortices  }

(see Fig.\ref{pairs of c-r vortices})
\end{figure}
\\

\hfill{}

\begin{centering}

4. REFERENCES

\end{centering} \hfill{}\\
 Davila, J. \& Hunt, J.C. R. Settling of particles near vortices and
in turbulence, \emph{Submitted to J. Fluid Mech.1999}. \vspace{1mm}\\
 Hyland, K. E., McKee, S. \& Reeks, M. W. 1999 Derivation of a kinetic
equation for the transport of particles in turbulent flows. \emph{J.
Phys. A: Math, Gen}., 6169-6190.\vspace{.75mm}\\
 Maxey, M. R. \& Corrsin, S. 1986 Gravitational settling of aerosol
particles in randomly oriented cellular flow fields. \textit{J. Atmos.
Sci} \textbf{43}, 1112-1134. \vspace{.75mm}\\
 Maxey, M. R., 1987 The gravitational settling of aerosol particles
in homogeneous turbulence and random flow fields. \emph{J. Fluid Mech}
\textbf{74}, 441-465.\vspace{.75mm}\\
 Pozorski, J. and Minier J-P. Probability density function modelling
of dispersed two-phase turbulent flows. \emph{Phys. Rev. E} \textbf{59},
1249-1261.\vspace{.75mm}\\
 Reeks, M. W. 1991 On a kinetic equation for the transport of particles
in turbulent flows\emph{. Phys. Fluids A} \textbf{3(3)}, 446-456.\vspace{.75mm}\\
 Reeks, M. W 1992 On the continuum equations for dispersed particle
flows in non-uniform flows\emph{. Phys. Fluids} 4, 1290-.\vspace{.75mm}\\
 Simonin, O. , Deutsch, E., and Minier, J. P. 1993 Eulerian prediction
of the fluid particle correlated motion in turbulent dispersed two-phase
flows. \textit{Appl. Sci. Res} \textbf{51}, 275-283.\vspace{.75mm}\\
 Swailes, D. C. and Darbyshire, K. F. 1997 \emph{Physics A} \textbf{242},
38- Wang, L-P and Maxey, M. R. 1993 Settling velocity and concentration
distribution of heavy particles in homogeneous isotropic turbulence.
\emph{J. Fluid Mech.} \textbf{256}, 27-68.\vspace{.75mm}\\
 Zaichik, L. I. \& Vinberg, A. A. 1991 Modelling of particle dynamics
and heat transfer in turbulent flows using equations for first and
second moments of the velocity and temperature fluctuations. Proc.
of the Eigth Symposium on Turbulent Shear Flows. Munich FRG Vol. 1,
pp. 1021-1026.

\hfill{}\hfill{}

\begin{centering}

APPENDIX A

\end{centering}

\hfill{} \hfill{}\\
 \emph{A1. Gaussian Process}\\
 \emph{\hfill{}}

We expand $\rho\left({\mathbf{r}-}\int_{0}^{t}\mathbf{v}_{p}(s)ds,0\right)$
in Eq.(\ref{eq.2forrho}) as a Taylor's series about $\rho({\mathbf{r}},0)$
so that formally 
\begin{equation}
\rho({\mathbf{r}},t)=exp\left\{ -\left[\int_{0}^{t}\nabla\cdot\mathbf{v}_{p}(s)ds+\int_{0}^{t}ds\,\mathbf{v}_{p}(s)\cdot\frac{\partial}{\partial{\mathbf{r}}}\right]\right\} \rho({\mathbf{r}},0).
\end{equation}
 We now suppose that both \-$\mathbf{v}_{p}(s)$ and $\nabla\cdot\mathbf{v}_{p}(s)$
to be a continuous processes whose statistics are correlated. That
is $\mathbf{v}_{p}(s)$ is the limit of the discrete process 
\begin{eqnarray}
\left[\mathbf{v}_{p}(s)\right] & \equiv & =_{N\rightarrow\infty}\left[\mathbf{v}_{p}(s_{1}),\mathbf{v}_{p}(s_{1})..\mathbf{v}_{p}(s_{j})..,\mathbf{v}_{p}(s_{N})\right]\nonumber \\
s_{j} & = & j\tau\,\,\,\textrm{with }N\tau=s.
\end{eqnarray}
 Similarly for $\nabla\cdot\mathbf{v}_{p}(s).$ For convenience we
specify a vector $\mathbf{q}(s)$
\begin{equation}
{\mathbf{q}}(s)=\left[{\textrm{v}}_{p1}\left(s\right),{\textrm{v}}_{p2}\left(s\right),{\textrm{v}}_{p3}\left(s\right),\nabla\cdot\mathbf{v}_{p}(s)\right]
\end{equation}
 whose statistics we specify through the characteristic functional
$M\left[\phi(s)\right]$ given formally by 
\begin{eqnarray}
M\left[{\phi(}s{)}\right] & = & \left\langle exp\left(i\int_{0}^{t}{\phi(}s{)\cdot\mathbf{q}(}s)ds\right)\right\rangle \nonumber \\
 &  & \textrm{where }\phi(s)\textrm{ is an arbitrary vector function of time}
\end{eqnarray}
 and we further assume that $\mathbf{q}(s)$ is Gaussian so that
\begin{eqnarray}
M\left[{\phi(s)}\right]= &  & exp\left\{ {i}\int_{0}^{t}\left\langle \mathbf{q}(s)\right\rangle \cdot\phi(s)ds\right.\nonumber \\
 &  & \left.-\frac{1}{2}\int_{0}^{t}ds_{1}\int_{0}^{t}ds_{2}\left\langle q_{i}^{\prime}(s_{1})q_{j}^{\prime}\left(s_{2}\right)\right\rangle \phi_{i}(s_{1})\phi_{j}(s_{2})\right\} \label{Gaussianfunctional}
\end{eqnarray}

where

\[
\begin{array}{ccc}
\left\{ \left\langle q_{i}{(}s)\right\rangle =W_{i}\,;\, q_{i}^{\prime}(s_{1})=q_{i}\left(s_{1}\right)-W_{i}\right\} \, i\leq3\,; & \left\{ \left\langle q_{i}(s)\right\rangle =0\,;\, q_{i}^{\prime}(s)=q_{i}(s)\right\} \, i=4\end{array}
\]

We recognize from the definition of the characteristic functional
that 
\begin{eqnarray}
\left\langle \rho({\mathbf{r}},t)\right\rangle  & = & M\left[i\phi(t)\right]\rho({\mathbf{r}},0)\nonumber \\
{\textrm{with }}\phi_{i}(t) & = & \frac{\partial}{\partial x_{i}}{\,\,\,\, for}i\leq3\nonumber \\
 & = & {\,\,1\,\,\,\,\,\,\,\,\,\,\,\,\,\,\,\,}i=4\label{eqnsforphi}
\end{eqnarray}
 and 
\begin{equation}
\left\langle \textrm{v}_{pi}(t)\rho({\mathbf{r}},t)\right\rangle =-\frac{\delta M\left[{i}{\phi}(s)\right]}{\delta\phi_{i}(t)}\rho({\mathbf{r}},0){for}i\leq3\label{functionaleqnforvn}
\end{equation}

\  $\bigskip$Substituting the Gaussian functional for $M$ given
in Eq.(\ref{Gaussianfunctional}) into Eq(\ref{functionaleqnforvn})
and performing the functional differentiation we obtain 
\begin{equation}
\left\langle \textrm{v}_{p_{k}}(t)\rho({\mathbf{r}},t)\right\rangle =\left\{ \left\langle q_{k}(t)\right\rangle +{i}\int_{0}^{t}\left\langle q_{4}(s)q_{k}(t)\right\rangle \phi_{4}(s)ds+{i}\sum_{i=1}^{3}\int_{0}^{t}ds\left\langle q_{i}(s)q_{k}(t)\right\rangle \phi_{i}(s)\right\} \left\langle \rho({\mathbf{r}},t)\right\rangle \label{nvintermsofq}
\end{equation}

\  $\bigskip$Substituting the values for $\phi_{i}(s)$ in Eq.(\ref{eqnsforphi})
into Eq.(\ref{nvintermsofq}) we obtain finally 
\begin{eqnarray}
\left\langle \textrm{v}_{p_{k}}(t)\rho({\mathbf{r}},t)\right\rangle =\left\{ \left\langle \textrm{v}_{p_{k}}{(\mathbf{r}},t)\right\rangle -\int_{0}^{t}\left\langle \textrm{v}_{p_{k}}(t)\,\nabla\cdot\textrm{v}_{p_{k}}(s)\right\rangle ds\right\} \left\langle \rho({\mathbf{r}},t)\right\rangle \nonumber \\
-\int_{0}^{t}ds{}\left\langle \textrm{v}_{p_{k}}^{\prime}(t)\textrm{v}_{p_{i}}^{\prime}(s)\right\rangle \frac{\partial}{\partial x_{i}}\left\langle \rho({\mathbf{r}},t)\right\rangle .
\end{eqnarray}

\hfill{} \hfill{}\\
 \emph{A1. Non-Gaussian Process}\\
 \emph{\hfill{}}

The same analysis can be extended to consider dispersion and drift
in which ${{\mathbf{v}}_{p}}{(}s)$ and ${{\nabla\cdot\mathbf{v}}_{p}}(s)$
are jointly non-Gaussian in which we express the characteristic functional
$M\left[{\phi}(t)\right]$ in terms of the cumulants of ${\mathbf{q}}\left(t\right),$i.e.
\begin{equation}
M\left[{\phi(}s{)}\right]=exp\left(\int_{0}^{t}\left\langle {\mathbf{q}}(s)\right\rangle {\cdot\phi(}s{)}ds{+}\sum_{m=2}^{\infty}\left[\begin{array}{c}
\frac{{i}^{n}}{m!}\int_{0}^{t}ds_{1}\int_{0}^{t}ds_{2}.....\int_{0}^{t}ds_{m}\times\\
\times\left\Vert \left\langle q_{i_{1}}^{\prime}(s_{1})q_{i_{2}}^{\prime}\left(s_{2}\right)....q_{i_{m}}^{\prime}\left(s_{m}\right)\right\rangle \right\Vert \times\\
\times\phi_{i_{1}}(s_{_{{i}}1})\phi_{i_{2}}\left(s_{2}\right)....\phi_{i_{m}}\left(s_{m}\right)
\end{array}\right]\right)
\end{equation}
 where $\left\Vert \left\langle q_{i_{1}}^{\prime}(s_{1})q_{i_{2}}^{\prime}\left(s_{2}\right)....q_{i_{m}}^{\prime}\left(s_{m}\right)\right\rangle \right\Vert $
represent the cumulants of ${\mathbf{q}(}t).$ Using this form for
$M\left[{\phi(}s{)}\right]$ and Eq.(\ref{functionaleqnforvn}) we
obtain: 
\begin{equation}
\left\langle {{\textrm{v}}_{p_{k}}}(t)\rho({\mathbf{r}},t)\right\rangle =\left\langle \rho({\mathbf{r}},t)\right\rangle W_{k}+\sum_{m=1}^{\infty}\left[\begin{array}{c}
\frac{\left(-1\right)^{m}}{m!}\int_{0}^{t}ds_{1}..\int_{0}^{t}ds_{m}\times\\
\left\Vert \left\langle q_{i_{1}}^{\prime}(s_{1})..q_{i_{m}}^{\prime}\left(s_{m}\right)q_{k}^{\prime}\left(t\right)\right\rangle \right\Vert \times\\
\phi_{i_{1}}(s_{1})..\phi_{i_{m}}\left(s_{m}\right)
\end{array}\right]\left\langle \rho({\mathbf{r}},t)\right\rangle \label{nvpnonGaussian}
\end{equation}
 with ${\phi}\left(t\right)$ given by Eqs(\ref{eqnsforphi}). So
picking out the contribution to the drift and to the gradient diffusion
from the correlation of $\,\,$\ the process $\left[{{\mathbf{v}}_{p}}{(}s)\right]$
with the process $\left[{{\nabla\cdot\mathbf{v}}_{p}}(s)\right]$
we can write Eqs(\ref{nvpnonGaussian}) more transparently as 
\begin{eqnarray}
\left\langle {{\textrm{v}}_{p_{k}}}(t)\rho({\mathbf{r}},t)\right\rangle  & = & \left\{ W_{k}-\sum_{m=1}^{\infty}\left(\begin{array}{c}
\frac{\left(-1\right)^{m+1}}{m!}\int_{0}^{t}ds_{1}..\int_{0}^{t}ds_{m}\times\\
\left\Vert \left\langle {{\nabla\cdot\mathbf{v}}_{p}}(s_{1})..{{\nabla\cdot\mathbf{v}}_{p}}(s_{m}){{\textrm{v}}_{p_{k}}}^{\prime}(t)\right\rangle \right\Vert 
\end{array}\right)\right\} \left\langle \rho(\mathbf{r},t)\right\rangle \\
 &  & \left\{ \begin{array}{c}
-\int_{0}^{t}ds_{1}\left\langle {{\textrm{v}}_{p_{i}}}^{\prime}(s_{1}){{\textrm{v}}_{p_{j}}}^{\prime}(t)\right\rangle +\\
\sum_{m=2}^{\infty}\left(\begin{array}{c}
\left(-1\right)^{m}\int_{0}^{t}ds_{1}...\int_{0}^{t}ds_{m}\times\\
\left\Vert \left\langle ..{{\textrm{v}}_{p_{i}}}^{\prime}(s_{1})..{{\nabla\cdot\mathbf{v}}_{p}}(s_{m}){{\textrm{v}}_{p_{k}}}^{\prime}(t)\right\rangle \right\Vert 
\end{array}\right)
\end{array}\right\} \frac{\partial\left\langle \rho(\mathbf{r},t)\right\rangle }{\partial x_{i}}+....
\end{eqnarray}

So to first order in the triple moments of $\left[{\mathbf{q}}(t)\right],$the
convective velocity ${\mathbf{v}}_{d}$ and Diffusion coefficients
$D_{ij}$ are respectively 
\begin{equation}
{\mathbf{v}}_{d}=\left\langle {\mathbf{v}}_{p}({\mathbf{r},\mathbf{t})}\right\rangle -\int_{0}^{t}ds_{1}\left\langle {{\nabla\cdot\mathbf{v}}_{p}}(s_{1}){{\mathbf{v}}_{p}}(t)\right\rangle +\frac{{\small1}}{{\small2}}\int_{0}^{t}ds_{1}\int_{0}^{t}ds_{2}\left\langle {{\nabla\cdot\mathbf{v}}_{p}}(s_{1}){{\nabla\cdot\mathbf{v}}_{p}}(s_{2}){{\mathbf{v}}_{p}}^{\prime}(t)\right\rangle +...\label{nonGaussiandriftvelocity}
\end{equation}
\begin{equation}
D_{ij}=\int_{0}^{t}ds_{1}\left\langle {{v}_{pi}}^{\prime}(s_{1}){{v}_{pj}}^{\prime}(t)\right\rangle -\int_{0}^{t}ds_{1}\int_{0}^{t}ds_{2}\left\langle {{\nabla\cdot\mathbf{v}}_{p}}(s_{1})..{{\textrm{v}}_{pi}}^{\prime}(s_{2}){{\textrm{v}}_{pj}}^{\prime}(t)\right\rangle 
\end{equation}
 So the diffusion coefficient is derived from two parts: one which
is appropriate for incompressible flows if the process and the other
which is appropriate for compressible flows for non-Gaussian processes.

\hfill{}\hfill{}

\begin{centering}

APPENDIX B Evaluation of \( \left\langle G\mathbf{u}(\mathbf{x},t)\right\rangle  \)

\end{centering}\hfill{}

We can formally write Eq.(\ref{<G>equation}) as 
\begin{equation}
\langle G\rangle=\left\langle exp\left\{ -\left[\mathbf{v}_{p}^{\prime}(t)\cdot\frac{\partial}{\partial\mathbf{v}}+\int_{t^{\prime}}^{t}ds\mathbf{v}_{p}^{\prime}(s)\cdot\frac{\partial}{\partial\mathbf{x}}+\int_{t^{\prime}}^{t}ds\nabla\cdot\mathbf{v}_{p}(s)\right]\right\} \right\rangle G^{(0)}(\mathbf{v},\mathbf{x},t|\mathbf{v}^{\prime},\mathbf{x}^{\prime},t^{\prime})
\end{equation}
 where $\mathbf{v}_{p}(s)$ is used as shorthand for $\mathbf{v}_{p}(\mathbf{v}^{\prime},t\prime|\mathbf{X}_{p}(\mathbf{x},t|s),s)$
and a similar short hand of $\nabla\cdot\mathbf{v}_{p}(s)$ for $\nabla\cdot\mathbf{v}_{p}(\mathbf{v}^{\prime},t\prime|\mathbf{X}_{p}(\mathbf{x},t|s),s)$.
and $\mathbf{v}_{p}^{\prime}(s)$ is the fluctuating value of $\mathbf{v}_{p}(s)$
with respect to its average value $\langle\mathbf{v}_{p}(s)\rangle$
. $G^{(0)}(\mathbf{v},\mathbf{x},t)$ is the response function 
\begin{equation}
G^{(0)}=\delta(\mathbf{v}-\langle\mathbf{v}_{p}(t)\rangle\delta(\mathbf{x}-\int_{t^{\prime}}^{\textrm{t}}ds\,(\left\langle \mathbf{v}_{p}(s)\right\rangle -\mathbf{x}^{\prime}))
\end{equation}
 So as for the passive scalar case we consider the statistical process
\begin{equation}
{\mathbf{q}}(s)=\left[{\textrm{v}}_{p1}^{\prime}\left(s\right),{\textrm{v}^{\prime}}_{p2}\left(s\right),{\textrm{v}^{\prime}}_{p3}\left(s\right),u_{1}^{\prime}(s),u_{2}^{\prime}(s),u_{3}^{\prime}(s),\nabla\cdot\mathbf{v}_{p}(s),\right]
\end{equation}
 with a given characteristic functional $M[\phi(s)]$ which we will
assume is a Gaussian functional. We have thus as before
\begin{eqnarray}
\langle G\rangle & = & M[\mathbf{i}\phi(s)]\\
\textrm{with }\phi_{i}(s) & = & \delta(s-t)\frac{\partial}{\partial\textrm{v}_{i}}+\frac{\partial}{\partial x_{i}}\textrm{ }\,\,\,\textrm{for }1\leq i\leq3\nonumber \\
 & = & 0\,\,\,\,\,\,\,\,\,\,\,\,\,\,\,\,\,\,\,\,\,\,\textrm{for }4\leq i\leq6\nonumber \\
 & = & 1\,\,\,\,\,\,\,\,\,\,\,\,\,\,\,\,\,\,\,\,\,\,\textrm{for }i=7
\end{eqnarray}

and 
\begin{equation}
\langle u_{i}(t)G\rangle=\frac{\delta M[\mathbf{i}\phi_{i+3}(s)]}{\delta\phi_{i+3}(t)}G^{(0)}(\mathbf{v},\mathbf{x},t|\mathbf{v}^{\prime},\mathbf{x}^{\prime},t)
\end{equation}
 Performing this functional differentiation on the Gaussian Characteristic
functional, leads to the closed expression
\begin{equation}
\langle u_{i}^{\prime}(t)G\rangle=-\left\{ \langle u_{i}^{\prime}(t)\textrm{v}_{p_{j}}^{\prime}(t)\rangle\frac{\partial}{\partial\textrm{v}_{j}}+\int_{t^{\prime}}^{t}ds\,\langle u_{i}^{\prime}(t)\textrm{v}_{p_{j}}^{\prime}(s)\rangle\frac{\partial}{\partial x_{j}}+\int_{t^{\prime}}^{t}ds\,\langle u_{i}^{\prime}(t)\nabla\cdot\mathbf{v}_{p}(s)\rangle\right\} \langle G\rangle\label{pdfclosureapprox.}
\end{equation}
 
\end{document}